\documentclass[]{raa}            
\usepackage{graphicx,times}
\usepackage{natbib}
\usepackage{epstopdf}

\providecommand{\bm}[1]{\mbox{\boldmath$#1$\unboldmath}}

\begin{document}

\newcommand{\beq}[1]{\begin{equation}\label{#1}}
\newcommand{\eeq}{\end{equation}}
\newcommand{\bea}{\begin{eqnarray}}
\newcommand{\eea}{\end{eqnarray}}
\newcommand{\gl}{COSMOS~J095930+023427}
\def\kmsmpc{km~s$^{-1}$~Mpc$^{-1}$}
\def\ho{H$_0$}
\newcommand{\degree}{\ensuremath{^\circ}}
\newcommand{\lens}{COSMOS~095930+023427}

 \def\disp{\displaystyle}

\title{A multi-wavelength study of the gravitational lens COSMOS~J095930+023427}

\volnopage{ {\bf 2012} Vol.\ {\bf XX} No. {\bf XX}, 000--000}
   \setcounter{page}{1}

\author{S. Cao\inst{1,2}, G. Covone\inst{2,3}, M. Paolillo \inst{2,3}, and Z.-H. Zhu \inst{1}}

\institute{Department of Astronomy, Beijing Normal University, Beijing 100875, China; {\it zhuzh@bnu.edu.cn}\\
     \and
   Dipartimento di Scienze Fisiche, Universit\`a di Napoli ''Federico II'',
   Complesso Universitario di Monte S. Angelo, via Cinthia, 80126 Napoli, Italy\\
\and
INFN, Sezione di Napoli, 
Complesso Universitario di Monte S. Angelo, 
via Cinthia, 80126 Napoli, Italy\\
%
\vs \no
   {\small Received [year] [month] [day]; accepted [year] [month] [day] }
}

\abstract{We present a multi-wavelength study of the gravitational lens 
COSMOS~J095930+023427 ($z_l=0.892$), together
with the associated galaxy group located at $z\sim0.7$ along the line of sight and
the lensed background galaxy. 
 The source redshift is currently unknown, but estimated to be at $z_s \sim 2$.
The analysis is based on the available public HST, Subaru, Chandra imaging data, and VLT
spectroscopy.
The lensing system is an early-type galaxy showing a strong [OII] emission line, and
produces 4 bright images of the distant background source.
It has an Einstein radius of $0.79''$, about 4 times large than the effective radius.
We perform a lensing analysis using both a Singular Isothermal Ellipsoid (SIE) and a
Peudo-Isothermal Elliptical Mass Distribution (PIEMD) for the lensing galaxy, and find that the final results
on the total mass, the dark matter (DM) fraction within the Einstein radius and the external shear due to a
foreground galaxy group are robust with respect of the choice of the parametric model and the
source redshift (yet unknown). 
We measure the luminous mass from the photometric data, and find the DM fraction within the Einstein radius $f_{\rm DM}$ to 
be between $0.71\pm 0.13$ and $0.79 \pm 0.15$, depending on the unknown source redshift.
Meanwhile, the non-null external shear found in our lensing models supports
the presence and structure of a galaxy group at $z\sim0.7$, and an independent measurement of the 0.5-2 keV X-ray
luminosity within 20'' around the X-ray centroid provides a group mass of $M=(3-10)\times 10^{13}$ M$_{\odot}$,
in good agreement with the previous estimate derived through weak lensing analysis.
Finally, by inverting the HST/ACS I$_{814}$ image with the lensing equation, 
we obtain the reconstructed image of the magnified source galaxy, which has a scale of about 3.3 kpc at $z_s=2$ (2.7 kpc at $z_s=4$)
and the typical disturbed disk-like appearance observed in low-mass star-forming galaxies at $z\sim3$.
However, deep, spatially resolved spectroscopic data for similar lensed sources are still required to 
detected the first stage of galaxy evolution, since the available spectrum shows no clear feature due to the background source.
}

\authorrunning{S. Cao, et al. }
\titlerunning{The gravitational lensing system COSMOS J095930+023427}
   \maketitle

\section{Introduction}
\label{sec:introduction}

Since the discovery of the first gravitational lens by \citet{Walsh79},
strong gravitational lensing (GL) has developed into an important astrophysical tool.
It is nowadays a primary technique to probe the dark matter (DM) spatial distribution in cosmic structures
(on scales from galaxies to galaxy clusters, e.g. \citet{Massey10}),
investigation of the DM properties and nature (e.g., \citet{Grin07,Moustakas09}),
to study in great detail the faint population of high-redshift galaxies (e.g., Richard et al. 2008, Monna \& Covone 2012),
and to provide a measurement of the Hubble constant free of the calibration in the
cosmic distance ladder (e.g., \citet{Schechter05}) or the cosmological
parameters via statistical analysis (see, e.g., \citet{Cao12a,Cao12b} and references therein).

Compared with other observational techniques to probe the radial mass profile, such as kinematics of stellar population \citep{cap06,van08} and the temperature of X-ray emitting gas \citep{hum06,chu08}, the advantage of GL lies in its ability to investigate the total mass within the Einstein radius
without any assumptions about the dynamical state of the system. Moreover, about two hundred new strong galaxy-galaxy lenses, in which the lensed sources are extended galaxies, have been discovered by recent large systematic surveys such as the SDSS survey (SLACS: \citet{Bolton06,Bolton08,Allam07}),
the CFHT-LS surveys (SL2S: \citet{Cabanac07}), and the COSMOS survey \citep{Faure08,Jackson08}.
These valuable samples can be used not only to study the DM in the lensing galaxies but also the properties of high-redshift ($z\sim2$)
star-forming galaxies acting as sources.

The insights into the galaxy structure are emerging from two aspects of works.
One is the joint statistical analysis for a large sample of galaxy-scale lenses.
The DM content in the central regions for early-type galaxies has been extensively
investigated by \citet{cap06,Tortora09}, who found that the central DM fraction is an increasing function of the total mass.
Recently, by combining strong and weak gravitational lensing, \citet{Lagattuta10} have measured the average mass properties of a sample of 41 strong gravitational lenses at moderate redshift, and found that changing the inner slope of the DM
profile by $\sim20\%$ will yield a $\sim30\%$ change in stellar mass-to-light ratio.
The other is mapping the mass distribution of individual galaxy-scale lens systems.
\citet{Covone09} analyzed the total mass and the DM fraction
within the Einstein radius for a $\sim L^*$ S0 galaxy at $z=0.4656$ as a function of the lensed source redshift, and obtained some results
supporting the above mentioned statistical study at lower galaxy mass scale.

Meanwhile, the properties and mass distribution of high-redshift galaxies
in the distant universe ($z> 2-3$) is still an almost completely uncharted territory.
A detailed exploration of more high$-z$ galaxies by means of lensing magnification would allow a great insight into the first stages of galaxy formation.

In this paper, we present an analysis of the gravitational lens \lens, an early-type lensing galaxy
at $z_l=0.8923\pm0.0007$ (COSMOS J095930+023427),
discovered in the Cosmic Evolution Survey (COSMOS, \citep{Scoville07}) field by \citet{Jackson08},
by means of a systematic, visual inspection of lensing candidates.
The aim of the present paper is to present a multi-wavelength
study of this galaxy-scale gravitational lens, focusing on probing  the DM distribution in the inner regions of
a $z\sim0.8$ elliptical galaxy and the properties of the lensed background source.

\begin{figure}
\begin{center}$
\begin{array}{cc}
\includegraphics[width=6cm]{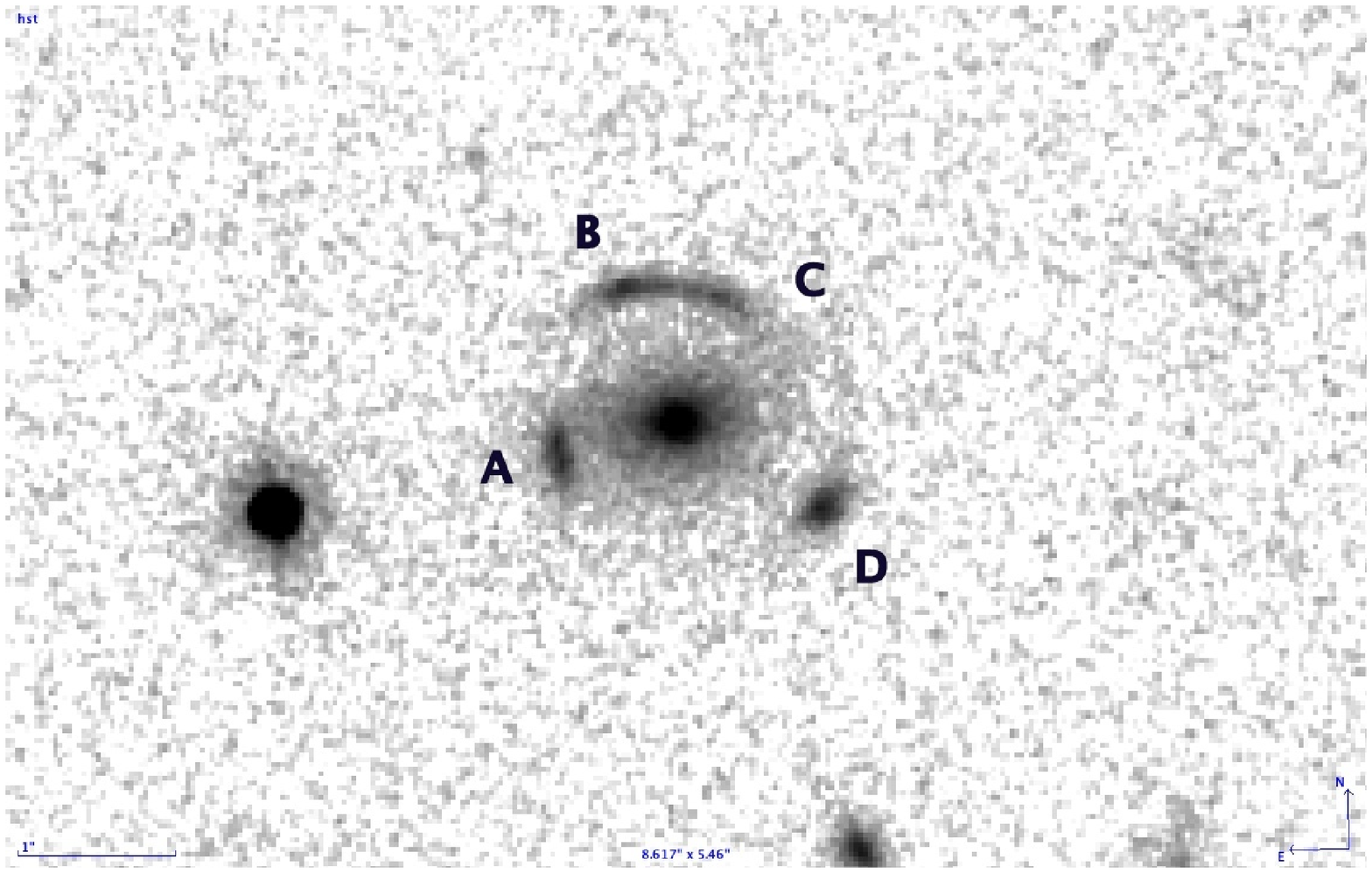}& 
\includegraphics[width=6cm]{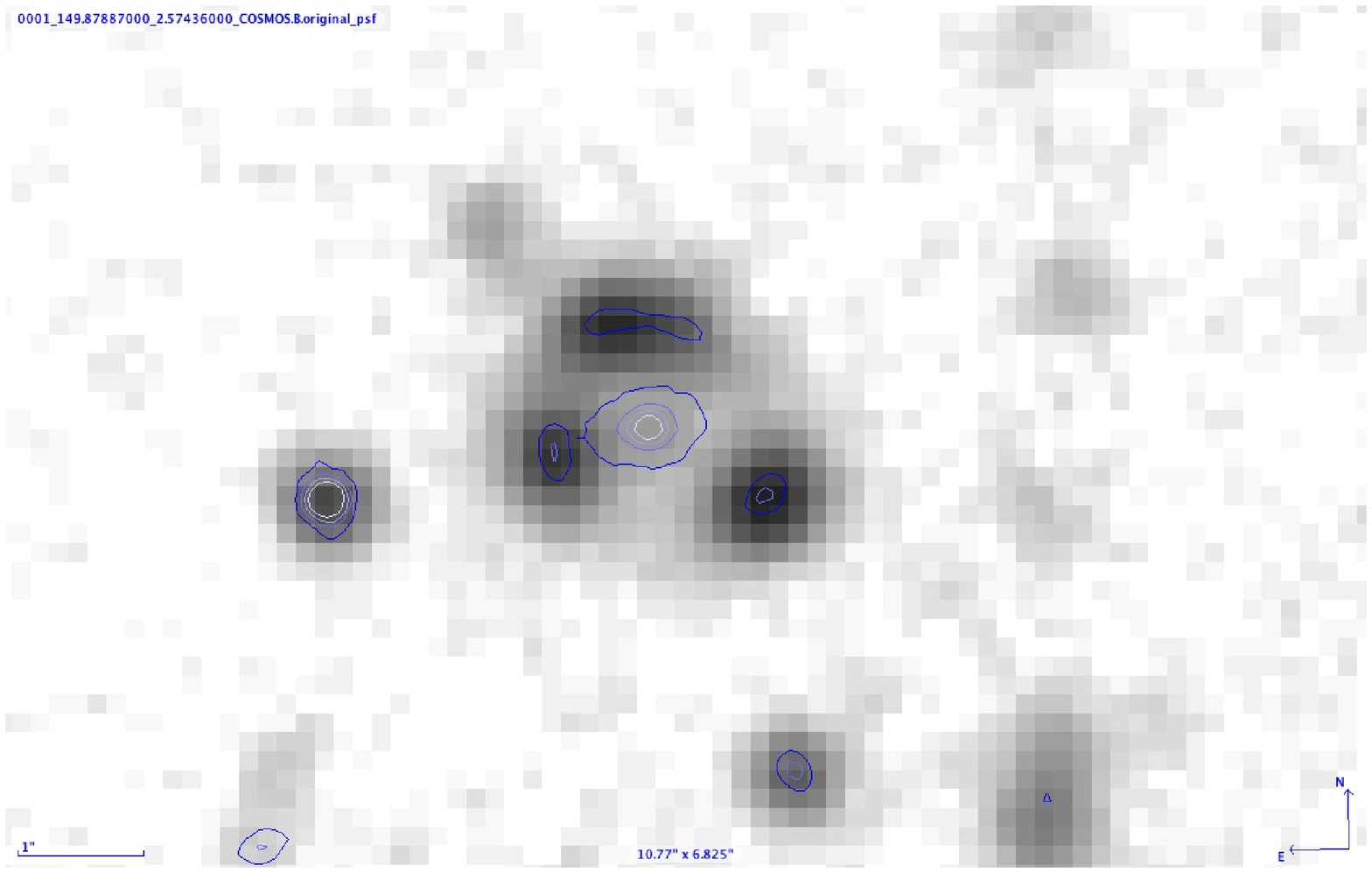} 
\end{array}$
\caption{Images of the gravitational lensing system
with the identification of the four lensed images.
Left panel: HST/ACS $I_{814}$ data.
Right panel: B-band SUBARU data, with contour of the sources from the HST/ACS data.
Field of view is $10'' \times 7''$. }
\label{fig:acs_subaru}
\end{center}
\end{figure}

The lensing system is clearly a four-image gravitational lens system (see Fig.~\ref{fig:acs_subaru}), with two merging images lying about 0\farcs8 
of the lensing galaxy and two images present to the southeast and southwest. The Subaru color imaging clearly shows that the lensed images have a similar blue colour, and the lensing galaxy dominates the light in the $I$ band, while hardly detected in the B image (Fig.~\ref{fig:acs_subaru}).

One important feature of J095930+023427 is that the lensing galaxy is bright and well separated from the lensed HST images,
which makes it possible to do a good determination of its photometric properties. Grasping information on both the mass and the light properties 
of this elliptical galaxy allows us to probe its DM fraction ($f_{\rm DM}$, defined as $f_{\rm DM}$=1-M$_\star$/(M$_\star$+M$_{\rm DM}$).
A study of the lensing galaxy was presented by \citet{Faure11}, who used a SIE model for the lensing galaxy and took into account the presence of a lower-redshift galaxy cluster along the line of sight. They obtained a DM fraction within the Einstein radius $f_{\rm DM} = 0.71^{+0.30}_{-0.12}$.

The goal of the present paper is to present a detailed multi-wavelength study of this lens,
including an improved lensing model based on the rich
data set obtained by the COSMOS collaboration for the lensing galaxy as well as a new insight of the mass distribution of the lensed young galaxy at $z\sim2$. This paper is organized as follows. In Section~\ref{sec:data}, we present HST imaging and Subaru spectroscopic observations of the the lens systems.
In Section~\ref{sec:model}, we introduce the strong-lens mass models and present the constraint results on the lensing galaxy. 
In Section~\ref{sec:stelar}, we briefly describe the light properties of a non-local elliptical galaxy and its corresponding DM fraction.
Section~\ref{sec:source} shows the main properties of the
reconstructed source galaxy. Discussion and conclusions appear in Section~\ref{sec:conclusions}.
Throughout this work, we assume a flat cosmology with
$\Omega_{\rm m}=0.3$, $\Omega_\Lambda=0.7$, $H_0=70$~km~s$^{-1}$~Mpc$^{-1}$.
The age of the Universe at the lens redshift is 6.2 Gyr.
All magnitudes presented in this paper are given in the AB system.

\section{The data} \label{sec:data}

In this section, we present the imaging and spectroscopic data used in the analysis of \lens.
The system was imaged as a part of the large observational effort to cover
the COSMOS  field. In our analysis, we use the imaging data from HST/ACS, Subaru/SuprimeCamera,
XMM, Chandra and spectroscopic data from zCOSMOS (Lilly et al. 2009).
A full description of the available dataset is given in \citet{Scoville07}, and we give here only a brief summary.

HST high-resolution images are essential to build an accurate  model of the gravitational lens,
while the large multi-wavelength coverage allows us to constrain the redshift and
the stellar population of the
lensing galaxy and the lensed source.
We obtain our data thorough the COSMOS Cutouts
service\footnote{Web site: \tt http://irsa.ipac.caltech.edu/data/COSMOS/index\_cutouts.html}.
HST/ACS images cover a square field of about 1.8 deg$^2$.
Final data have $0.03''$ per pixel scale, 
with depth $I_{814} \simeq 28$ and $\sim$50\% completeness for sources 0.5'' in diameter at $I_{814}=26.0$ mag.
Subaru data are obtained in six broad-band (B, $g'$, V, $r' $, $i'$, $z'$ ) and one narrow-band (NB816) filters \citep{Tani07}.
The Subaru color imaging clearly shows that the lensed images have a similar blue color and the lensing galaxy dominates the light in the $I$ band.

The lensing galaxy was selected among the spectroscopic targets of the zCOSMOS project \citep{Lilly09}.
We retrieved the spectra from the second data release,
containing the zCOSMOS-bright spectroscopic observations that are carried out using VLT/VIMOS.
The measured redshift is $z_l=0.8923\pm0.0007$, classified as probable redshift in \citet{Lilly09}.
The redshift measurement is based mostly on the robust identification of the [OII] emission line and a weak CaII-K absorption feature.
Unfortunately, the VLT spectrum shows no signature from the background source: as discussed hereafter, we can  rely 
only on the available photometric data to constrain its redshift.

The main arc (merging images B and C) is located at about $0.85''$ from the main galaxy center.
A first model of this system has been presented by Faure et al. (2011): they estimated the  Einstein radius 
to be $R_{\rm E} = 0.79\pm 0.02''$ \citep{Faure11}.  This
corresponds to a linear scale of 6 pkc: 
in the following, we constrain the total mass and 
dark matter fraction  of the galaxy, using the Faure et al. value of $R_{\rm E} $ in order to compare our results
on the same length scale.

\section{The lensing model}\label{sec:model}

In this section we introduce two gravitational lens models used to
measure the total mass of the lensing galaxy within R$_E$, M($<$R$_E$),
and to determine the intrinsic properties of the lensed source.
We adopt two different lens models: a Singular Isothermal Ellipsoid (SIE) and a Peudo-Isothermal Elliptical Mass Distribution (PIEMD) \citep{Kassiola93}.

The previous lensing-based studies have demonstrated that
the mass density profile of early-type galaxies is consistent with an
isothermal mass model, both in the inner regions (within the Einstein radii, see e.g. \citet{Lagattuta10,Ruff11})
and at distances of $\sim 50-300 ~h^{-1}$ kpc \citep{she04,man06}.
Therefore we choose not to constrain the density profile and adopt  a SIE model following the normalization proposed by \citet{kormann94}. 
The ellipticity $\varepsilon$ is related to the axis ratio $q = \sqrt{\frac{1-\varepsilon}{1+\varepsilon}}$ \citep{Keeton98}. 
We have therefore three free parameters: the central velocity dispersion ($\sigma$), the shape and orientation of the total mass distribution.

Following recent works (e.g., \citet{Jullo07}), we also model the galaxy with a PIEMD, 
which is widely used for strong lensing studies both
on galaxy scale and galaxy clusters scale \citep{Covone06,Donnarumma09}.
Five free parameters are included in this model:
the core radius ($r_0$), cut-off radius ($r_c$), and velocity dispersion ($\sigma$), the shape and orientation of the total mass distribution.
As $r_c$ vanishes, the potential approaches a singular isothermal potential, truncated at the cut-off radius.

The main lensing galaxy is clearly not isolated: 
two massive early-type galaxies are found within $\sim 20''$, see Fig.~\ref{fig:SIE}. 
These two galaxies are likely members of a galaxy group located along the line of sight at $z\sim0.7$
(see Sect~\ref{sect:group}), that has already been considered in \citet{Faure11} as a part of the whole lensing system.
Therefore, in this section, we consider the contribution of these two massive galaxies on the lensing potential. 
For simplicity, we have modeled the two additional galaxies as singular isothermal spheres (SIS),
with the velocity dispersion as the only free parameter.
However, we do not assume {\it a priori} the presence of the galaxy group, that can be revealed through the presence of an external shear.
Such an external shear allows to characterize the contribution of the
environment to the total lens potential. Therefore, we consider two additional lensing models
with two more free parameters: the external shear parametrized by a shear strength $\gamma$
and its orientation PA$_\gamma$ (two models SIE+SIS+SIS+$\gamma$ and PIEMD+SIS+SIS+$\gamma$, respectively).

Optimization of the lens models has been performed by means of 
Lenstool \citep{Kneib93, Jullo07}, a tool adopting a Bayesian approach to strong lensing modelling.
The best model is obtained by  
reproducing the location of the observed multiple images (i.e., positions of the brightest peak and shape)
within the supplied uncertainties. 
We also checked that supplying the expected position of the critical curve between images B and C does not significantly  improve the results.
We choose to use the source plane minimization algorithm for higher accuracy.
In Lenstool, the positions of the images are the main input observational data to constrain the lens model parameters. In our analysis,
the position of the brightest peak is determined as the image position, and the error on each image position is assumed to be $0.05''$.
In Table~\ref{table1} we provide the lens galaxy central coordinates as well as the positions of the multiple
images used in the lens modeling. A detailed discussion of the lensing models is given below.
We assume the source to be located at redshifts $z=2$ and $4$, when deriving the total mass measurements.

\begin{table}
\centering
\caption{Positions of the lensing galaxy and the images (in degree) used as constraints for the lens mass model. \label{table1}}
\begin{tabular}{c c  c  }
\hline
\hline
Lens &RA&DEC\\
{\it  Image}& {\it  RA }&{\it DEC}\\
\hline
COSMOS~J095930+023427& 149.878930& 2.574342\\
{\it  A} &{\it 149.87902}&{\it  2.5745776} \\
{\it  B} &{\it 149.87886 }&{\it 2.5745646 }\\
{\it C} &{\it 149.87915} &{\it 2.5742854 } \\
{\it D }&{\it 149.87869} &{\it 2.5741922} \\
\hline
\end{tabular}
\end{table}

\begin{table*}
\centering
\caption{\label{constraint} Best fit parameters for the lens models, assuming $z_s=2$:
(1) Model name. 
(2) Bayesian evidence ${\rm log} E$. 
(3) $\chi^2$ of the best lens model.
(4) Shear parameters.
(5) Best-fit parameters for the main galaxy.
(6) Velocity dispersion of the galaxy \#2. 
(7) Velocity dispersion of the  galaxy \#3.
(8) Mass of the main galaxy in the Einstein radius.
(9) DM fraction within the Einstein radius.}
\begin{tabular}{c c c c  c  c  c  c c}
\hline
\hline
Model    & ${\rm log} E$ &$\chi^2$  & ($\gamma$, PA$_\gamma$) &($\sigma_1$,$\varepsilon$,PA)& $\sigma_2$ & $\sigma_3$ & M($<$R$_E$)& $f_{DM}$($<$R$_E$)\\
               &  &              &   &     ({\rm km~s$^{-1}$}) & {\rm km~s$^{-1}$} &{\rm km~s$^{-1}$} & {\rm 10$^{11}$M$_\odot$}& \\
\hline
SIE+SIS+SIS  &  -22.06 & 1.7  & \_   &(238,0.28,-10$\degree)$ &   391 & 603 & 3.49 & 0.81$^{+0.15}_{-0.15}$\\
SIE+SIS+SIS$+\gamma$  & -21.04 & 2.6  & (0.25,+60$\degree)$  &(231,0.28,+0.52$\degree)$ &388 &600 & 3.50 & 0.81$^{+0.15}_{-0.15}$\\
\hline
Model   & ${\rm log} E$  &$\chi^2$  & ($\gamma$, PA$_\gamma$) &($\sigma_1$,$r_0$,$r_c$)& $\sigma_2$ & $\sigma_3$ & M($<$R$_E$)& $f_{DM}$($<$R$_E$)\\
                 & &              &   &     ({\rm km~s$^{-1}$}, {\rm kpc}, {\rm kpc}) & {\rm km~s$^{-1}$} &{\rm km~s$^{-1}$} & {\rm 10$^{11}$M$_\odot$}& \\
\hline
PIEMD+SIS+SIS &  -22.30 & 1.8   & \_  &(200,0.14,24) &372 & 596 & 3.25 & 0.79$^{+0.12}_{-0.12}$\\
PIEMD+SIS+SIS$+\gamma$ & -18.29 & 5.6   &  (0.24,+31$\degree$)  &(199,0.21,37) &373 &596 & 3.24 & 0.79$^{+0.12}_{-0.12}$\\
\hline
\end{tabular}
\end{table*}

\subsection{SIE+SIS+SIS} \label{SIE}

In order to asses the effect of the external shear produced by the group and cluster,
we have dealt with two simple models, namely the SIE+SIS+SIS and the SIE+SIS+SIS$+\gamma$.

The main galaxy is described by a SIE, defined by its position, velocity dispersion ($\sigma_1$), 
orientation (PA) and ellipticity ($\varepsilon$).
The SIE central position is fixed to that of the light centroid for
the lensing galaxy J095930+023427, as determined in the HST/ACS 
image.
Moreover, considering that there might be a possible misalignment between the luminous bulge and the 
DM halo \citep{Kochanek02}, an 
additional $\pm$20$\degree$ uncertainty is placed on the orientation of the Sersic bulge light profile. 
The prior of the SIE ellipticity is $\varepsilon$=[0.0,0.9], which take into account a possibly shallower distribution of DM 
halo \citep{Gavazzi07}. For the second and third galaxies in the model,
we fix the SIS central position to their position in the I-band catalog and the only varying parameter 
is the velocity dispersion ($\sigma_2 \, ,\sigma_3$). The external shear is parametrized by 
two free parameters including a shear strength ($\gamma$)
and orientation (PA$_\gamma$) with the following priors: $\gamma$=[0.0,0.5], PA$_\gamma$=[-90$\degree$,90$\degree$].
Therefore, the total number of free parameters for the two models is 5 and 7, respectively.

The best-fit values of the model parameters and the values of 
 corresponding Bayesian evidence E and $\chi^2$
(assuming a source located at redshift $z_s=2$) are reported in Table~\ref{constraint}.
While the quantity ${\rm log } E$ can effectively be used to rank lensing models, see Jullo et al. (2007),
we report them mainly to show the overall relative quality of the fitting models.

\begin{figure*}
\begin{center}
\includegraphics[width=6cm]{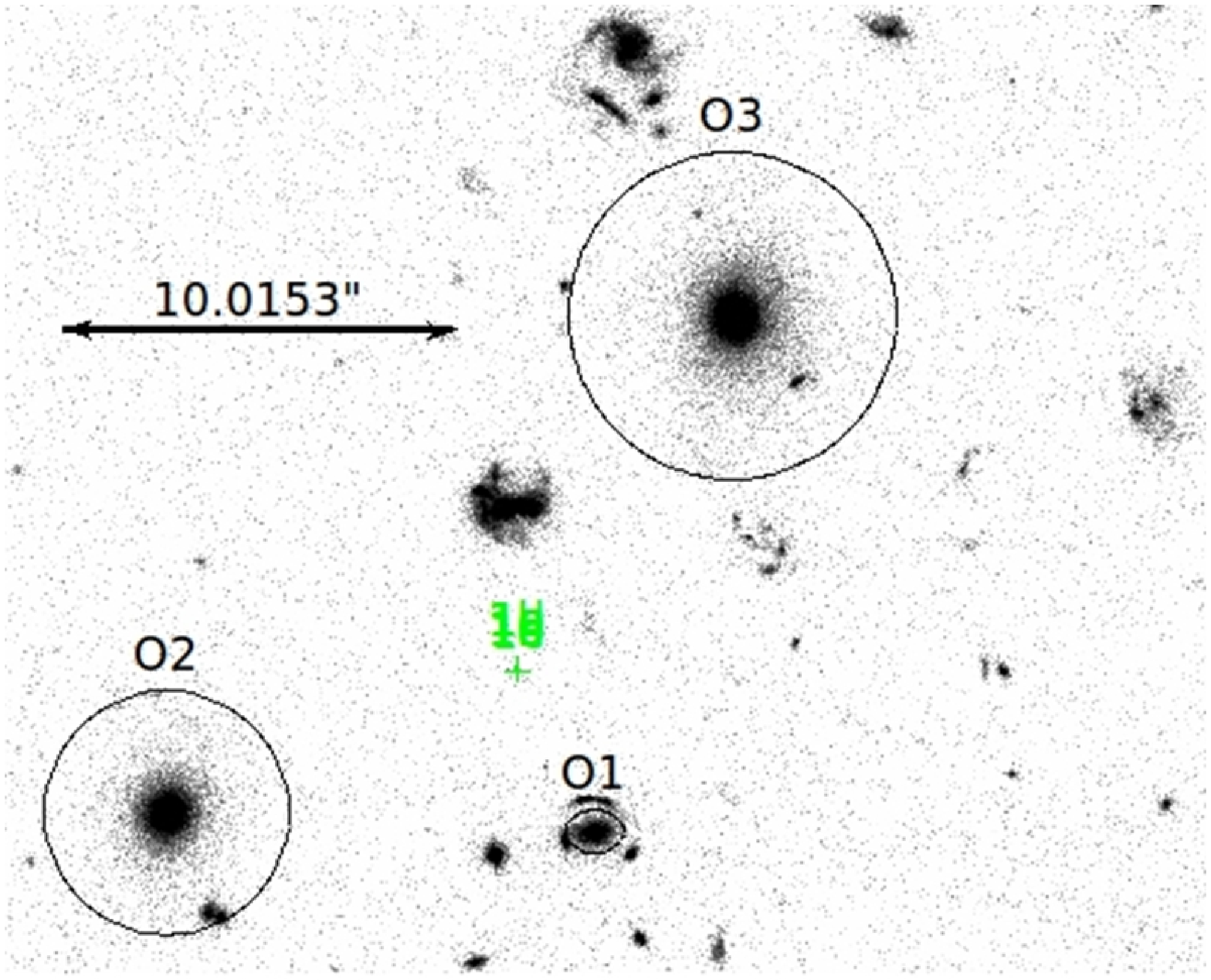}
\includegraphics[width=6cm]{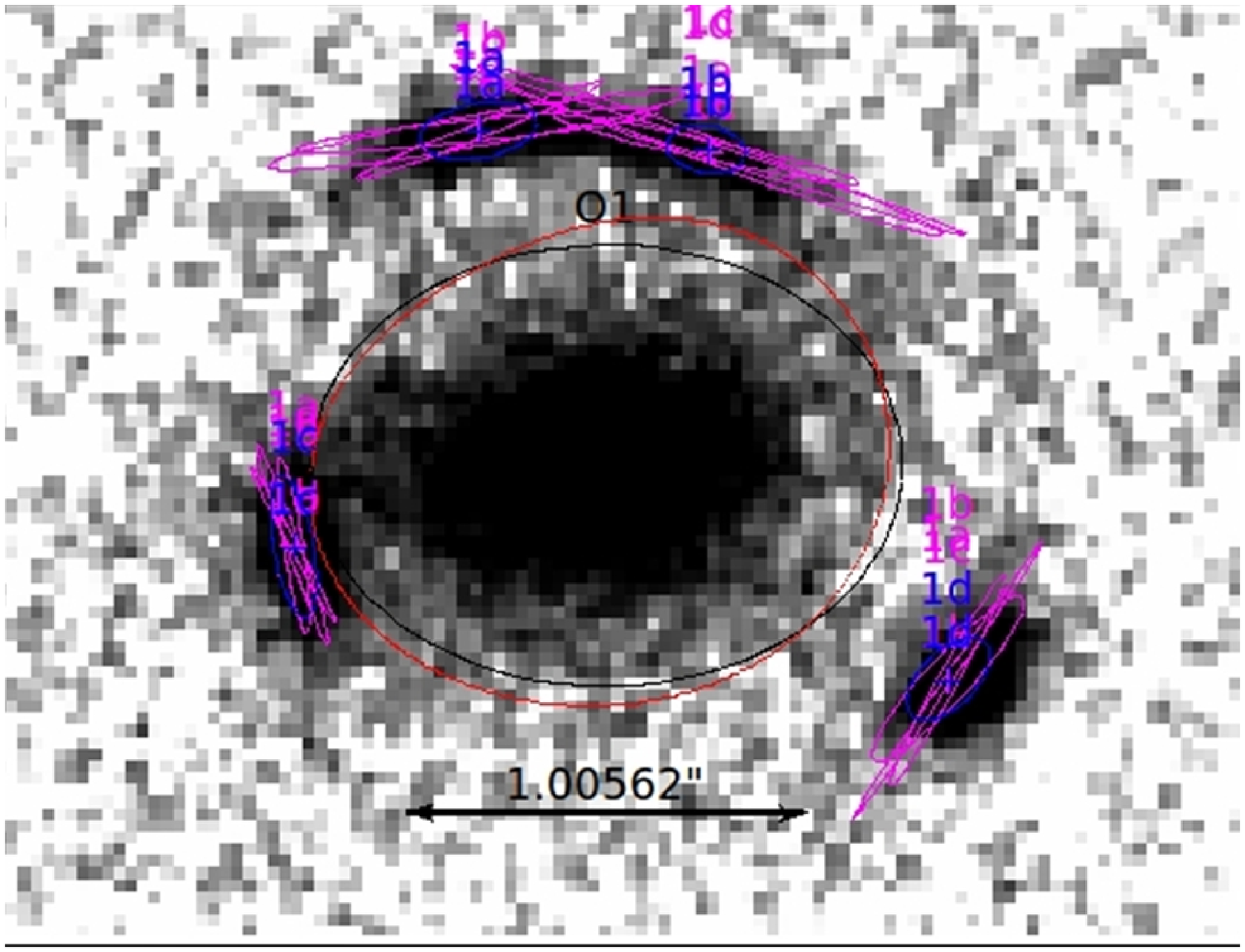} 
\caption{ Mass models with the SIE+SIS+SIS$+\gamma$ model.
Left panel: Constraint results with the black circles representing the main galaxy and the secondary and third galaxy 
(the favorite source positions are denoted by green ellipses).
Right panel: zoom on the lens galaxy and on the images. North is to the top and East to the left.}
\label{fig:SIE}
\end{center}
\end{figure*}

We also recomputed the best-fit models for a source at redshift $z_s=4$, in order to
estimate systematic errors on the DM fractions, as discussed below.
From the best-fit model displayed in Fig.~\ref{fig:SIE},
we see that the fit is (at face value) better when compared with the results of \citet{Faure11}:
with additional observational constraints such as the velocity dispersions of the lens galaxy closest neighbors,
it would be possible to obtain a more detailed model of the lens potential which would most probably improve the fit.

The image positions are slightly better reproduced (in terms of $\chi^2$)
when an external shear is included in the models (when using both the SIE and PIEMD),
as generally expected when both internal and external sources of anisotropy are included in the model.
However, the value of the shear strength $\gamma$ is quite significant since it amounts to about $\gamma\sim0.2$.
Therefore, despite the large shear strength, we shall call this latter model as the fiducial lens model in the following,
with the model parameter constraint results displayed in Table~\ref{constraint} and Fig.~\ref{fig:SIE}.
The total mass of the main galaxy in the Einstein radius would be
M($<$R$_E$)=3.5$^{+0.5}_{-0.3} \times$10$^{11}$M$_\odot$, a result compatible with that of \citet{Faure11}.
We stress here that the velocity dispersion associated with the secondary and third lens can be associated with the galaxy mass,
however, we choose not to discuss their masses in this paper, due to the simplifying assumption on the corresponding potential SIS.
An improved version of the present mass model will require future measurements of their velocity dispersions.
In neither case (with or without external shear), a significative offset between the mass distribution in the lensing model (i.e., the total mass distribution) 
and and the light distribution is found. Indeed, we obtain a value of the ellipticity of the total mass distribution ($\epsilon=0.28$) that is coherent
with the ellipticity of the light distribution ($\epsilon=0.21 \pm 0.05$, see \citet{Faure11}).

Finally, the predicted source positions (for each of the 4 elliptical-like
images) are given by our fiducial model in Fig.~\ref{fig:SIE}, and are clearly overlapping with each other.
Further constraints and discussion on this lensed galaxy are given in Sect.~\ref{sec:source}.

\begin{figure*}
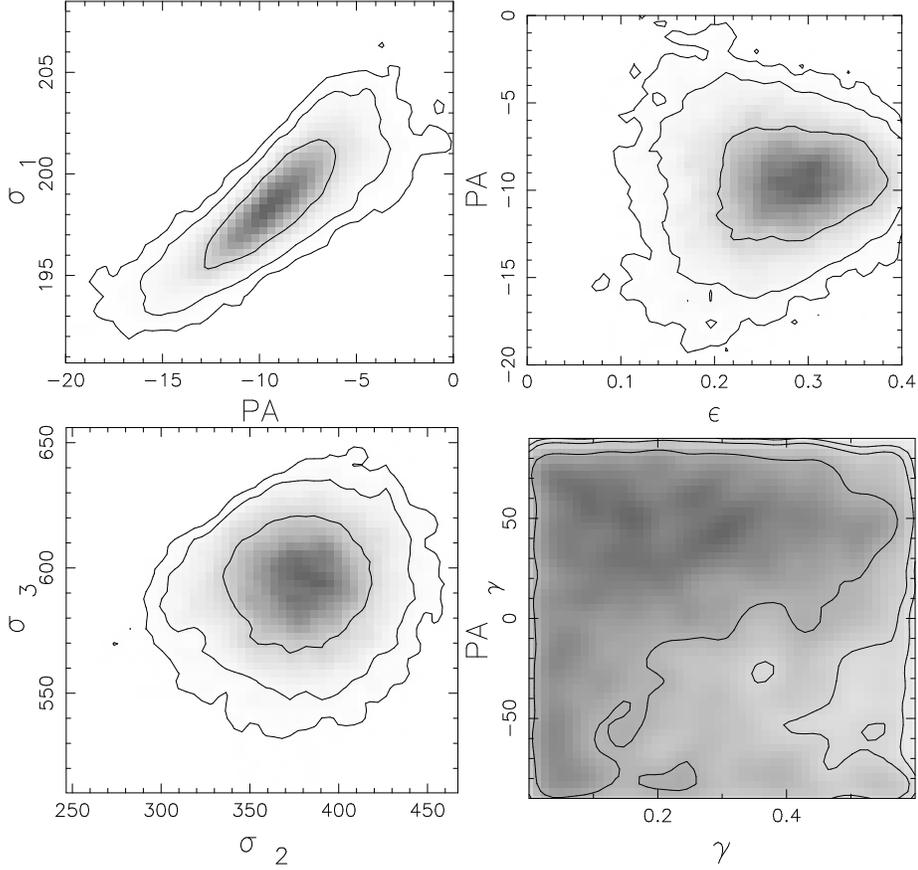

\begin{center}
\includegraphics[width=6cm]{ms1158fig5.eps} 
\includegraphics[width=6cm]{ms1158fig6.eps} 
\includegraphics[width=6cm]{ms1158fig7.eps} 
\includegraphics[width=6cm]{ms1158fig8.eps} 
\caption{2D marginalized PDFs for the parameters of the PIEMD+SIS+SIS+$\gamma$ model.
Top panels: velocity dispersion of the main galaxy vs galaxy PA; galaxy PA vs ellipticity.
Bottom panels: velocity dispersion of the galaxy \#2 and \#3; external shear vs shear angle.
The three contours stand for the 68\%, 95\% and 99.5\% confidence levels.}
\end{center}
\label{fig:PIEMD}
\end{figure*}

\subsection{PIEMD+SIS+SIS} \label{PIEMD}

The lens galaxy model and corresponding mass within Einstein radius are
shown in Table~\ref{SIE}, with M($<$R$_E$)=3.25$\times$10$^{11}$M$_\odot$.
The PIEMD+SIS+SIS+$\gamma$ model yields a similar fit in terms of Bayesian evidence and  $\chi^2$,
with a relatively large external shear ($\gamma=0.24$).
This clearly supports complementary evidence for the presence of
a massive galaxy group located along the line of sight.
Also using this parametric model, we measure an ellipticity the total mass distribution ($\epsilon=0.22$)
in perfect agreement with that of the light distribution: the DM halo does not appear to be shallower
than the stellar distribution.

In Fig.~\ref{fig:PIEMD} we plot the 2D marginalized probability distribution functions (PDFs)
for the parameters of the PIEMD+SIS+SIS+$\gamma$ model.
We note that the large value for the velocity dispersion of the most massive galaxy in the lensing system
($\sigma_3 \sim 600 \, {\rm km/s}$) can be due to the contribution of the surrounding galaxy groups.
In the bottom right panel we show the PDF for the shear parameters: it is evident that
these parameters are not as strongly contained as the other model parameters.
While the best-fit value $\gamma = 0.24$ is quite large when compared with
estimates of the shear produced by the cosmic environments of strong lensing galaxies
($\gamma$ ranging from 0.02 to 0.17, \citep{Wong11}),
very low vales for the shear $\gamma$ are still well within the $1\sigma$ contour. 
A further study of the present lensing system requires taking into account the detailed galaxy distribution along the line-of-sight.

\subsection{The DM fraction}

In this section, we determine the stellar mass within the Einstein radius for the main galaxy, M$_{\star}$($<$R$_E$) and
compare it with the total mass within the Einstein radius, M($<$R$_E$) obtained from lens modeling.
Correspondingly, the projected DM fraction within the Einstein radius: 
$f_{DM}$($<$R$_E$)= 1-$\frac{{\rm M}_\star{\rm (<R}_E)}{{\rm M(<R}_E)}$ is obtained.
These values are reported in Table~\ref{constraint} when assuming a source located at $z_s=2$.

The stellar mass is derived following \citet{Ilbert10} and assuming a Salpeter initial mass function (IMF) from 0.1 to 100 $M_{\odot}$.
The total stellar mass is found to be $M_* = (0.91 \pm 0.23) \times 10^{11} \, M_{\odot}$. Hence,  
the stellar mass inside the Einstein radius can be derived by measuring the light profile, and considering the fraction of the total light inside that radius.
We considered here for the slight lensing magnification ($\simeq 1.2$) of the galaxy flux by the nearby galaxy group.
We obtain $ M_* (R_{\rm E})= 0.68 \pm 0.18 \times 10^{11} \, M_{\odot} $,  slightly smaller than the value found by  \citet{Faure11}.
We can now use this to determine the DM fraction inside $R_{\rm E}$ and for the whole galaxy.
By using the best-fit values given in Table~\ref{constraint}
and considering a source located at redshift $z_s=2$, we obtain a value $f_{\rm DM} = 0.79 \pm 0.12$ (with the PIEMD model).
The major source of uncertainty in our losing model is given by the unknown source redshift.
When considering a more distant source at $z_s=4$, we obtain a less massive lensing galaxy:
the total mass within the Einstein radius is then measured to be
$M (R_{\rm E})  = 2.5 \, \times 10^{11} \,  M_{\odot} $
and $2.3 \, \times 10^{11} \,  M_{\odot} $, with the SIE and PIEMD models, respectively, regardless of the external shear.
This translates in a lower allowed value for the DM fraction: $f_{\rm DM} = 0.82 \pm 0.12$.
While considering the whole galaxy, the DM fraction is $0.95 \pm 0.10$.
Our measurements at $\sim 4$ effective radii can be directly compared with the findings
by \citet{Deason12}, studying a sample of 15 nearby elliptical galaxies, by using planetary nebulae and globular clusters as
kinematical tracers.
Within 4~$R_{\rm eff}$, the DM fraction is found to be between 0.6 and 0.9, for the galaxy mass range considered here.
A numerous sample at high redshift is required in order to detect any evolution inn the DM fraction,
as a function of galaxy mass.

\section{The properties of the galaxy group} \label{sect:group}

\begin{figure}
\centering
\includegraphics[width=10cm]{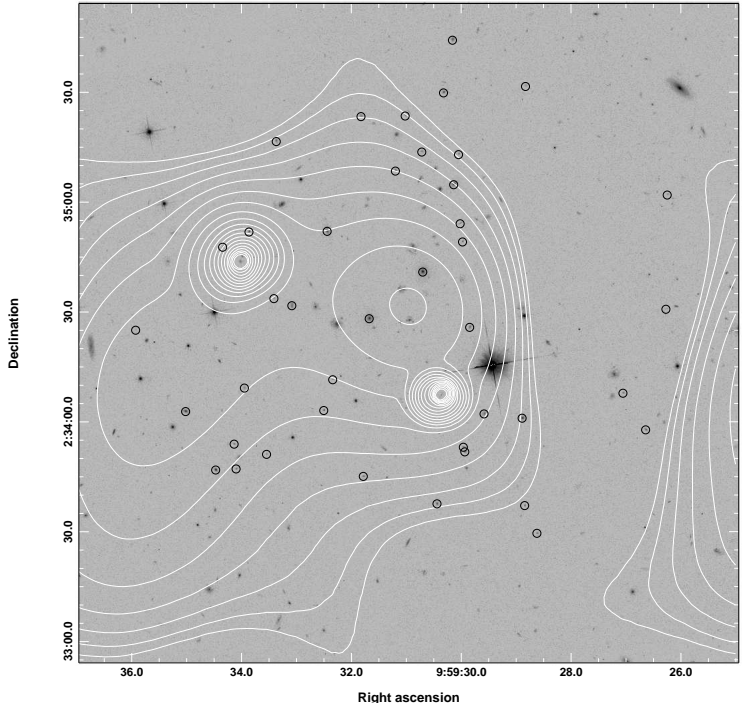} 
\caption{HST/ACS I band image of the region around the lens galaxy.
The white contours show the X-ray surface brightness distribution in the 0.5-2 keV band,
as derived through adaptive smoothing of the raw X-ray image.
Black circles mark the position of the candidate member galaxies of the group,
according to the analysis of George et al. (2011).}
\label{group288}
\end{figure}

As discussed above, a significative external shear in the lensing model points to
the presence of massive structure along the line of sight
to the lensing system. This adds up with other, several strong evidences for the presence of a galaxy group at $z \sim 0.5$:
a spatially extended X-ray emission, and an over-density in the photometric redshift distribution,
confirmed by the more sparse spectroscopic data.
Recently, George et al. (2011) have presented a spectroscopic survey
of galaxy groups in the  redshift range $0<z<1$,  including this structure.


The analysis of the COSMOS field performed by \citet{Finoguenov07} allows to identify galaxy clusters and groups based
on their X-ray emission. The galaxy membership to those groups was assessed in \citet{George11}, based on the distance
from the X-ray centroid, as well as photometric and spectroscopic redshifts.

Based on these analysis, \citet{Faure11} noticed that there is a galaxy group (ID~288 according to the classification in the X-ray and galaxy catalogs) along the line of sight toward our lens, likely at $z\sim0.7$.
We use the COSMOS public data\footnote{http://irsa.ipac.caltech.edu/data/COSMOS/} to verify the properties of such group of galaxies.
In Fig.~\ref{group288} we show the HST/ACS I band image of the field around our lens; we overlay on the optical image,
the X-ray contours obtained through adaptive smoothing of the 0.5-2 keV Chandra observations of the same region.
The adaptive smoothing procedure allows to separate the emission of the two bright AGNs present in the field, from the
spatially extended emission likely associated with the galaxy group.
Using the most recent versions of the X-ray and galaxy
catalogs \footnote{The catalogs are available at http://astro.berkeley.edu/~mgeorge/cosmos/},
we identified the galaxies which are candidate group members, which are shown by
black circles in Fig.~\ref{group288}.

Although the X-ray detection has a low significance
($\sim 2\sigma$ from our adaptively smoothed map, and quality flag 2 in the X-ray catalog), the emission appears to be
centered between the two brightest group galaxies, and aligned with the main gravitational lens.
While the X-ray contours suggests the presence of an eastern tail, which overlaps with the optical group members, the significance of this feature is too low to
decide whether it is an artifact of the smoothing process or it is related to the sparse distribution of the group galaxies.

To make an independent assessment of the group mass, we measure the 0.5-2 keV X-ray
luminosity within 20'' (corresponding to a linear scale of 140 kpc at $z=0.7$) around the X-ray centroid, deriving a
$L_X=(8.6-13)\times 10^{42}$ erg/s for temperatures in the range $1-2$ keV and $Z/Z_{\odot}=0.6-1$.
Using the scaling relations provided by \citet{Eckmiller11}, this translates in a mass $M=(3-10)\times 10^{13}$ M$_{\odot}$, which is in good agreement with the estimate of $4.4\times 10^{13}$ M$_{\odot}$ derived by \citet{Leauthaud11} through weak lensing analysis.

\section{The galaxy stellar population}\label{sec:stelar}

\begin{figure*}
\begin{center}
\includegraphics[width=10cm]{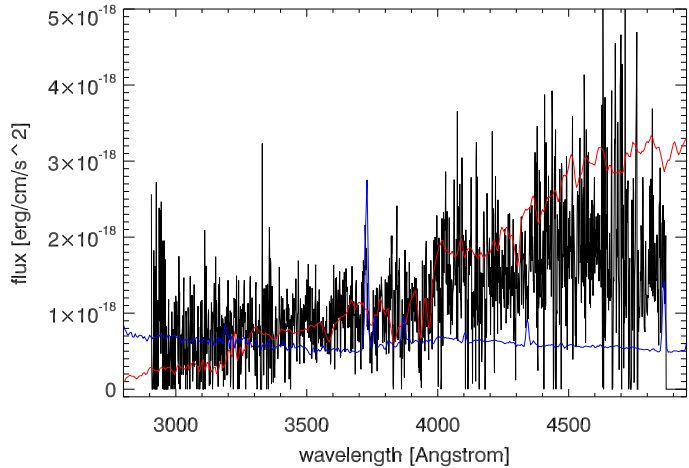} 
\caption{The optical spectrum of the lensing galaxy from zCOSMOS project (in black).
The [OII] emission line is evident.
The observed spectrum s compared with two spectral templates from \citet{Kinney96}: an elliptical galaxy (red)
and a star-forming galaxy (blue).}
\label{fig:spec}
\end{center}
\end{figure*}

The lensing galaxy is classified as an early-type galaxy by its morphology, and
the basic morphological parameters can be found in the
COSMOS Morphology Catalog (v1.1; \citet{Cassata07}).
The Sersic index is $n= 1.98\pm 0.20 $,
and the effective radius is $\theta_{\rm eff} = 0''.21 \pm0.08 \simeq 3.8 \, \theta_{\rm E}$.
However, the galaxy also shows an evident O[II] emission line, see Fig.~\ref{fig:spec}.
We measure its luminosity to be
$L_{\rm O[II]} \, = (8.1 \pm 5.9) \, \times 10^{40} \, {\rm erg} \, {\rm s}^{-1} \, .$
If interpreting this as a signature of ongoing star-formation, rather than AGN nuclear activity,
we can determine the galaxy star-formation rate (SFR) by
adopting the calibration introduced by \citet{Kennicutt98}:
\begin{equation}
{\rm SFR } \, (M_{\odot} / {\rm yr}^{-1}) \, = \,  1.4 \times 10^{-41} \, L(O[II]) \,   .
\end{equation}
We find therefore a modest SFR$\sim  1.2 M_{\odot} {\rm yr}^{-1}$.

Then, we have performed an analysis of the properties of the lensing galaxy stellar population
by using the available low-resolution spectrum from zCosmos and the tool STARLIGHT \citep{cid05,cid07b}.
We assume as a base a set of simple stellar populations with a Chabrier (2003)
initial mass function, and exclude the wavelength range of the observed emission line.
The measured (light-weighted) age of the lensing galaxy is $ \sim 3.4 $ Gyr, with the metallicity  $Z/Z_0 = 0.014$
($\chi^2 = 1.5$). Therefore, the galaxy formation epoch is located at redshift $z\sim2$: however, we caution that this
value is to be interpreted as a lower value, since
a likely young stellar population in the galaxy is biasing this measurement.

\section{Image reconstruction in the source plane}\label{sec:source}

Finally, as a demonstration of the capability of the present lensing system as a
{\it gravitational telescope}, we use the lensing model with the lower $\chi^2$ to invert the lensing equation and reconstruct the intrinsic source morphology.
With this aim, we used the task {\tt cleanlens} provided by Lenstool \citep{Keren12}:
for each point of the image plane, the program computes the corresponding point in the source plane.

\begin{figure*}
\begin{center}
\includegraphics[width=10cm]{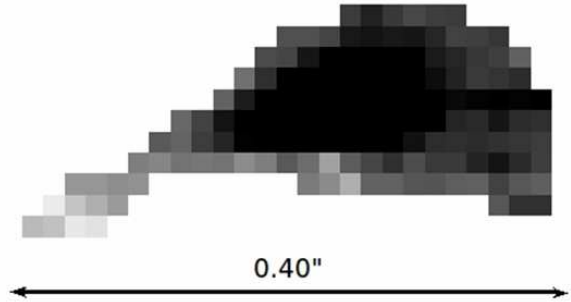} 
\caption{Image reconstruction of the lensed galaxy, based on the HST/ACS I$_{814}$ data.}
\label{fig:source}
\end{center}
\end{figure*}

The inversion result of the HST/ACS I$_{814}$ image is presented in Fig.~\ref{fig:source}.
The reconstructed source image allows us to investigate the detailed spatial structure of the magnified source. We find that the galaxy has a scale of about 3.3 kpc at $z_s=2$ (2.7 kpc at $z=4$), and a disturbed disk-like appearance which is typical in low-mass star-forming galaxies at $z\sim3$ \citep{Genzel11}.
Unfortunately, the available spectrum shows no clear feature due to the background source (see Fig.~\ref{fig:spec})
and only the HST/ACS image allows a clean separation of the lensed images from the deflector. Therefore deep, spatially resolved spectroscopic data are still required for a large sample of similar lensed sources (magnification $\sim10$, $z\sim3$)
to gain further insight into the first stage of galaxy evolution.

\section{Conclusions}\label{sec:conclusions}

We have performed a multi-wavelength analysis of the gravitational
lens COSMOS~J095930+023427 ($z_l=0.892$), based
on the large set of public data from the COSMOS survey.
The lensing system is an early-type galaxy with an Einstein radius of $0.79"$. We perform a lensing analysis using both a SIE and a PIEMD:
final results on the total mass, the dark matter (DM) fraction within the Einstein radius and the external shear due to a
foreground galaxy group are robust with respect of the choice of the parametric model and the
source redshift (yet unknown). We find that the DM fraction is $f_{\rm DM}=0.75$ at about 4 effective radii with a source at $z_s=2$,
similar to what recently found in local elliptical galaxies \citep{Deason12}.
The non-null external shear found in our lensing models supports
other observational data about the presence and structure of a galaxy group at $z\sim0.7$, which appears to be
centered between the two brightest group galaxies, and aligned with the main gravitational lens according to the X-ray detection.
An independent measurement of the 0.5-2 keV X-ray luminosity within 20'' around the X-ray centroid provides us a group mass of $M=(3-10)\times 10^{13}$ M$_{\odot}$, a result consistent with the previous estimate derived through weak lensing analysis.
We also find that the lensing galaxy shows a strong [OII] emission line, probably due to a low amount of ongoing star-formation activity. Finally, by inverting the lensing equation of the HST/ACS I$_{814}$ image, we obtain the reconstructed source image presented in Fig.~\ref{fig:source} and investigate the detailed spatial structure of the magnified source. The lensed galaxy has a scale of about 3.3 kpc at $z_s=2$ (2.7 kpc at $z_s=4$), and a typical disturbed disk-like appearance in low-mass star-forming galaxies at $z\sim3$ \citep{Genzel11}. However, we note that deep, spatially resolved spectroscopic data for similar lensed sources are still required to detected the first stage of galaxy evolution, since the available spectrum shows no clear feature due to the background source (see Fig.~\ref{fig:spec}).

The present multiwavelength study is meant to be the first step in a systematic, individual
study of the large sample of gravitational lenses found in the COSMOS field \citep{Faure08,Jackson08},
with the aim to probe the evolution of the structure of the inner regions of elliptical galaxies.
A companion spatially-resolved spectroscopic observational campaign is the most natural step to complement the
dataset already in hand and obtain a deeper insight into the properties of massive early-type galaxies
at $z \sim 0.7 - 1.0$.

\begin{acknowledgements}
We acknowledge fruitful discussions with E. Jullo and L. Izzo.
This work is supported by the National Natural Science Foundation
of China under the Distinguished Young Scholar Grant 10825313 and
Grant 11073005, the Ministry of Science and Technology national
basic science Program (Project 973) under Grant No. 2012CB821804, the
Fundamental Research Funds for the Central Universities and Scientific Research Foundation of Beijing Normal University,
and the Excellent Doctoral Dissertation of Beijing Normal University Engagement Fund.

\end{acknowledgements}

\end{document}